\documentclass[prd,preprint,showpacs,groupedaddress]{revtex4-1}
\usepackage{amsmath}
\usepackage{amsfonts}
\usepackage{amssymb}
\usepackage{geometry}
\usepackage{natbib}
\usepackage[english]{babel}
\usepackage{graphicx}
\usepackage{epstopdf}
\usepackage{subfigure}
\usepackage{caption}
\usepackage{multirow}
\usepackage{indentfirst}
\usepackage{mathrsfs}
\usepackage[colorlinks,linkcolor=blue,anchorcolor=blue,citecolor=blue]{hyperref}

\begin{document}
		
\setlength{\parindent}{2em}
\title{Bardeen black hole in magnetically charged four-dimensional Einstein-Gauss-Bonnet gravity}
\author {Shu-Jun Zhang} \author{He-Xu Zhang} \author{Lei Shao} \author{Jian-Bo Deng} \author{Xian-Ru Hu} \email[Xian-Ru Hu:]{huxianru@lzu.edu.cn}
\affiliation{Institute of Theoretical Physics $\&$ Research Center of Gravitation, Lanzhou University, Lanzhou 730000, China}
		
\begin{abstract}
In this paper, we investigate the shadow radius and quasinormal modes of a four-dimensional magnetically charged Einstein-Gauss-Bonnet Bardeen black hole and point out a simple connection between them in the eikonal limit. By studying a massless scalar field perturbation in this spacetime background and using the sixth-order Wentzel–Kramers–Brillouin (WKB) approximation, we get the quasinormal modes(QNMs) and perform a detailed analysis. It shows that the quasinormal modes are depend on the Gauss-Bonnet coupling constant $\alpha$ and the magnetic charge $g$. When the value of $g$ increases, the values of the real parts of the QNMs increase and those of the imaginary parts decrease. There is a similar effect on $\alpha$. We also give a formula of the QNMs and shadow radius in the eikonal limit and check it numerically for the real and imaginary part of it, respectively.
\end{abstract}
		
\maketitle
\section{Introduction}  
Physicists are always fascinated by Black hole physics and pay attention to it over decades. This is not only because that black hole physics itself is deserved to be well studied, but also theoretically it has deeply connection to many other different subjects of physics, especially in string theory and AdS/CFT \cite{GHSS,CMS,KPSR,DHVB,JMM,DTSAOS}. Nowadays, we have much better understanding of black holes in practical since the detection of gravitational waves (GWs) of black hole binary mergers \cite{BPAT} by the LIGO and VIRGO observatories and the captured image of the black hole shadow of a supermassive M87 black hole by the Event Horizon Telescope Collaboration \cite{EHT1,EHT4}. 
\par
In general, shadows and quasinormal modes(QNMs) are important features of black holes. Shadow is a dark region in a bright background, which is formed by the black hole's accretion of photons and strong gravitational lensing effect. Researches have found the expressions of shadow radius of black holes and have well studied the influence of black hole about parameters on the shapes and sizes of the shadows \cite{JLS,JPL,MGL,KSGG,ZCMHD1,ZCMHD2,PYFA}. QNMs, which are asymptotic solutions of perturbation fields around compact objects under certain boundary conditions, are complex frequencies. The real parts represent oscillation and the imaginary parts are proportional to the decay rate of a given mode. A perturbed black hole emits GWs exactly in the form of quasinormal radiation and we can detect them by LIGO etc. The perturbation theory of Schwarzschild black hole and its stability under small perturbations was first studied by Regge, Wheeler \cite{RW} and Zerilli \cite{FJZ}. After that, QNMs have been investigated by analytic and numerical methods \cite{BBM,BFSWILL,EWL1,EWL2,KDKBFS,HPNO,VCJPSL,RAK5,EBKDK,JLJ,JLJQYP,VCMBWZ,SYG,CCDHN,JMMO,WLWLQ,TTAS,PR,DSEMR,XCCM,MSCS}. By the way, QNMs are also appeared in the quantization of the black hole \cite{SHOD,OD,GKUN,RAKAZO}.
\par
In this paper, we investigate the shadow radius and QNMs of the four-dimensional magnetically charged Einstein-Gauss-Bonnet Bardeen black hole. Shadow radius is closely related to QNMs. It was Cardoso $\emph{et al.}$ \cite{VCMBWZ} first showed that in the eikonal limit, the real part of the QNMs was related to the angular velocity of the last circular, null geodesic, and the imaginary part was related to the Lyapunov exponent. Then Stefanov $\emph{et al.}$ \cite{RAKAFZ} found a connection between geodesics and quasinormal modes in eikonal limit. Cuadros-Meglar \cite{BCMFO} found an analytical relation between the eikonal limit of quasinormal frequencies and the black hole shadow. Jusufi \cite{KJUS1,KJUS2} pointed out that in the eikonal regime the real part of QNMs is inverse proportional to the shadow radius. He explored the effect of the matter parameter $k$ on the QNMs of massless scalar field and electromagnetic field perturbations in a black hole spacetime surrounded by perfect fluid dark matter and showed out that there exists a reflecting point $k_0$ corresponding to maximal values for the real part of QNMs frequencies.Then Chen \emph{et al.} \cite{CDY} investigated four-dimensional Einstein-Gauss-Bonnet charged Bardeen black hole and verified such correspondence between the shadow and the scalar test field. It is remarkable that this correspondence does not hold in general, such as for gravitational filed \cite{RAKZS}. With these results in mind, we shall explore the connection between the QNMs and the shadow radius. We can use the WKB approach to compute the QNMs frequencies. This approach was used by Schutz and Will \cite{SW}, developed to the third order by Iyer and Will \cite{IW}. In the present paper, we shall use the sixth order WKB approximation for calculating QNMs developed by Konoplya \cite{RAK}. The four-dimensional black hole solutions in Einstein-Gauss-Bonnet gravity have been a challenge, while for higher dimensions the solutions have been obtained. Recently, a four-dimensional theory of gravity with Gauss-Bonnet correction is introduced by the re-scaling of Gauss-Bonnet coupling $\alpha \rightarrow \frac{\alpha}{D-4}$ and taking the limit $(D-4) \rightarrow 0$ \cite{GL}.The spherically symmetric black hole solutions are also obtained in the same paper. The generalization to other black holes has also appeared \cite{AGM,PGSF,WL,HLYP,SGGRK,BEPKJH,KJU3,LMAHL}.
\par
The rest is organized as follows. In the next section, we investigate the photon sphere and shadow radius of the four-dimensional charged Einstein-Gauss-Bonnet black hole. In Section III, we calculate the QNMs by the $6th$ order WKB approximation method and the effects of the Gauss-Bonnet coupling constant and the magnetic charge are discussed. We also study the connection between the QNMs and shadow radius, Lyapunov exponent. In the last section our conclusions are presented.

\section{The photon sphere and shadow radius of Bardeen black hole in 4D EGB gravity}\label{II}
Consider the action for four-dimensional Einstein-Gauss-Bonnet gravity with nonlinear electrodynamics
\begin{equation}
S=\frac{1}{16\pi}\int d^4x \sqrt{-g}[R+\alpha(R^2-4R_{\mu\nu}R^{\mu\nu}+R_{\mu\nu\rho\sigma}R^{\mu\nu\rho\sigma})-4\mathcal{L}(F)],
\label{eq2.1}
\end{equation}
where $\alpha$ is the Gauss-Bonnet coupling constant, $\mathcal{L}(F)$ is a function of $F\equiv \frac{1}{4} F_{\mu\nu}F^{\mu\nu}$ and $F_{\mu\nu} \equiv 2\nabla_{[\mu}A_{\nu]}$ is the electromagnetic field tensor. More specifically, $\mathcal{L}(F)$ is given by \cite{ABEAG1,ABEAG2}
\begin{equation}
	\mathcal{L}(F)=\frac{3}{2sg^2}\left [\frac{\sqrt{2g^2F}}{1+\sqrt{2g^2F}} \right ]^{\frac{5}{2}},
	\label{eq2.2}
\end{equation}
where $s \equiv \frac{|g|}{2M}$, and $g$ and $M$ are the magnetic charge and the mass of the magnetic monopole.
\par 
The Bardeen balck hole solution, which is the static and spherically symmetric solution of the four-dimentional Einstein-Gauss-Bonnet gravity with nonlinear electrodynamics, is given by
\begin{equation}
	ds^2 = -f(r)dt^2 + \frac{1}{f(r)}dr^2 + r^2 (d\theta^2+\sin^2\theta d\phi^2),
	\label{eq2.3}
\end{equation}
where
\begin{equation}
	f(r)= 1+\frac{r^2}{2\alpha} \left  [1 \pm \sqrt{1+\frac{8 M \alpha }{(r^2+g^2)^{3/2}}}  \right].
	\label{eq2.4}
\end{equation}
\par
As we can see, there are two branches of the solution,``$+/- $" represents ``positive/negative" branch, respectively. In the limit $\alpha \rightarrow 0$ and far region condition, the negative branch recovers to the Bardeen black hole with positive gravitational mass, whereas the positive branch is unstable due to the negative gravitational mass \cite{BGDDS}. Furthermore, in the limit $g \rightarrow 0$ the negative branch recovers to the Gauss-Bonnet Schwarzchild solution. Therefore, we shall only consider the negative branch of the solution in the rest of the paper.
\par
To obtain the horizon of the black hole,we just let $f(r) \equiv 0$,
\begin{equation}
	1+\frac{r^2}{2\alpha} \left [1\pm \sqrt{1+\frac{8 M \alpha }{(r^2+g^2)^{3/2}}} \right]=0.
	\label{eq2.5}
\end{equation}
This equation can be solved numerically for suitable values of the parameters. Generally, there are two solutions $r_{\pm}$, where $r_+$ is the event horizon we focus on and $r_-$is the inner horizon. For convenience, we let $M=1$ throughout this paper.
\par 
To obtain the null geodesic in above black hole spacetime, we studied the Lagrangian of a free photon orbiting around the equatorial orbit of the black hole

\begin{equation}
	\mathcal{L} = \frac{1}{2}\left[-f(r)\dot{t}^2 + \frac{1}{f(r)}\dot{r}^2 + r^2\dot{\phi}^2\right],
	\label{eq2.6}
\end{equation}
where the dot represents the differentiation with respect to an affine parameter. Then, the generalized momenta are given by

\begin{eqnarray}
	p_t = -f(r)\dot{t}=-E,\\
	p_r = \frac{\dot{r}}{f(r)},\\
	p_{\phi} = r^2\dot{\phi}=L,
	\label{eq2.7}
\end{eqnarray}
 where $E$ and $L$ as conserved quantities are constants and represent the energy and orbital angular momentum of the photon, since there are two Killing vector fields $\frac{\partial}{\partial t}$ and $\frac{\partial}{\partial \phi}$ in this spacetime. Then, the Hamiltonian \cite{WL2} is 

\begin{eqnarray}
	\mathcal{H} = p_{\mu}\dot{x}^{\mu}- \mathcal{L} = \frac{1}{2}(-E\dot{t} +\frac{\dot{r}^2}{f(r)} +L\dot{\phi})=0.
	\label{eq2.8}
\end{eqnarray}

From above We obtained the equation of radial geodesic in terms of the effective potential $V_{eff}$ 

\begin{eqnarray}
	\dot{r}^2 +V_{eff}(r)=0, \\
	V_{eff}(r)=- E^2+ \frac{L^2}{r^2}f(r).
	\label{eq2.9}
\end{eqnarray}

To investigate the geometric sharp of the shadow of the black hole, one needs to solve the unstable orbit condition for the photon
\begin{eqnarray}
	V_{eff}(r)=0, \quad\quad \frac{\partial V_{eff}(r)}{\partial r} = 0, \quad\quad \frac{\partial^2 V_{eff}(r)}{\partial r^2} <0.
	\label{eq2.10}
\end{eqnarray}

Solving the second equation of the above equations,

\begin{equation}
	2-\frac{rf'(r)}{f(r)}=0.
	\label{eq2.11}
\end{equation}

From this equation, one can obtain the radius of the photon sphere $r_{ps}$.
\par 

\begin{table}[htbp]
	\begin{center}
		\begin{tabular}
			{|c|c|c|c|c|c|c|c|c|}
			\hline
			$g$&
			$\alpha $&
			$ R_{sh} $&
			$g$&
			$\alpha $&
			$ R_{sh} $&
			$g$&
			$\alpha $&
			$ R_{sh} $\\
			\hline
			\multirow{5}*{0.1}&
			-0.1&
			5.22557&
			\multirow{5}*{0.3}&
			-0.1&
			5.15674&
			\multirow{5}*{0.5}&
			-0.1&
			5.00788 \\
			&
			-0.3&
			5.29794&
			&
			-0.3&
			5.23373& 
			&
			-0.3&
			5.09654\\
			&
			-0.5&
			5.36589&
			&
			-0.5&
			5.30551&
			&
			-0.5&
			5.17767\\
			&
			-0.7&
			5.43009&
			&
			-0.7&
			5.37295&
			&			
			-0.7&
			5.25282 \\
			&
			-0.9&
			5.49105&
			&
			-0.9&
			5.43670&
			&
			-0.9&
			5.32308 \\
			\hline
		\end{tabular}
	\end{center}  
	\label{tab1}
	Table I. The size of shadow radius for different values of $g,\alpha$.
\end{table}

Furthermore, it is found that the shadow radius of the black holes $R_{sh}$ have a simple relation with the radius of the photon sphere $r_{ps}$,

\begin{equation}
	R_{sh}=\left. \frac{r}{\sqrt{f(r)}} \right |_{r=r_{ps}}.
	\label{eq2.12}
\end{equation} 

By solving Eqs. (\ref{eq2.4}), (\ref{eq2.11}), and (\ref{eq2.12}), we would get the expression for the radius of the photon sphere and the shadow radius. But for our metric the specific expressions of them are very complex and are not presented here and we gave numerical results of shadow radius for different values of $g$ and $\alpha$. They are listed in Table I. One should notice that $\alpha$ cannot be too negative, because the metric function may not be real inside the event horizon when $\alpha$ is too negative \cite{TCPCFM,MGPCL,CYZPCLG}.
\par 
From Table I, we found that the  shadow radius is closely related to the magnetic charge $g$ and Gauss-Bonnet coupling constant $\alpha$. When $g$ is fixed and the value of $\alpha$ decreases, the shadow radius $R_{sh}$ increases. There is a similar relation between $g$ and $R_{sh}$, namely, when $\alpha$ is fixed and $g$ increases, $R_{sh}$ decreases.

\section{QNMs of the scalar field}
As an important feature of black holes, QNMs are one of important ways to detect black holes. As a particular example, let us consider a  massless scalar field perturbation in the metric (\ref{eq2.3}), which is described by the following equation
\begin{equation}
\frac{1}{\sqrt{-g}}\partial_{\mu}\left(\sqrt{-g}g^{\mu\nu}\partial_{\nu}\Phi(t,r,\theta,\phi)\right)=0.
\label{eq3.1}
\end{equation}

\par
By separating variables, the function $\Phi(t,r,\theta,\phi)$ for the
scalar field can be given in terms of the spherical harmonics
\begin{equation}
	\Phi=\sum\limits_{l,m}{e^{-i\omega t}\Phi(r)r^{-1}Y_{lm}(\theta,\phi)},
	\label{eq3.2}
\end{equation}
where $\omega$ is the perturbation frequency and $e^{-i\omega t}$ represents the time evolution of the scalar field. Putting Eq. (\ref{eq3.2}) into Eq. (\ref{eq3.1}), one can show that the field perturbation equation in the black hole spacetime is given by the Schrödinger wave-like equation,
\begin{equation}
	\frac{d^2 \Psi}{d\partial_{r_{\star}}^2}+ (\omega^2 - \mathcal{V}(r))\Psi(r)=0,
	\label{eq3.3}
\end{equation}
where 
\begin{equation}
	\mathcal{V}(r)=f(r) \left [\frac{f^{\prime}(r)}{r}+\frac{l(l+1)}{r^2} \right ],
	\label{eq3.4}
\end{equation}
here we have employed the "tortoise" coordinate
\begin{equation}
    dr_{\star}=\frac{dr}{f(r)}.
    \label{eq3.5}
\end{equation}

\par
 We adopted following boundary conditions to make the real part of QNMs positive,
\begin{equation}
	\Phi(r_{\star})\sim \exp[\pm i\omega r_{\star}],\qquad r \rightarrow \pm \infty,
	\label{eq3.6}
\end{equation}
where $\omega$ can be further written in terms of the real and imaginary parts, i.e., $\omega=\omega_R-i\omega_I$, where $\omega_I$ is also positive here. The real and imaginary parts represent the oscillation frequency and the decay rate, respectively. 
\par 
With the equation for the effective potential, we can use the WKB approach to compute the QNMs frequencies. In the present paper, we shall use the 6th WKB approximation \cite{RAK}, which developed by Iyer and Will \cite{SW}, for calculating QNMs. We calculated values of quasinormal modes for the scalar perturbations with various magnetic charges and Gauss-Bonnet coupling constants given in Table II. In Table III, we provided the values of QNMs having $l= 1, 2, 3$ and $n= 0, 1$, respectively. We also draw corresponding figures to Tables II and III, respectively. Here we did not calculate the modes with $l=n=0$ since this approach does not work well in $l \leq n$. However, one can use the Frobenius method for more accurate calculation of this fundamental mode \cite{RAKZSAZ}.

\begin{table}[htbp]
	\begin{center}
		\begin{tabular}{|c|c|c|c|c|}
			\hline
			&$g=0.1$ &
			$g=0.3$&
			 $g=0.5$&
			 $g=0.8$ \\
			\hline
			$\alpha$=-0.1&
			0.291025-0.099073i&
			0.295166-0.098291i&
			0.304427-0.096130i&
			0.335212-0.082764i \\
			\hline
            $ \alpha$ =-0.2&
            0.288556-0.100505i&
            0.292512-0.099937i&
            0.301333-0.098346i&
            0.330725-0.089183i \\
            \hline
            $ \alpha$ =-0.3&
            0.285964-0.102003i&
            0.289742-0.101635i&
            0.298144-0.100550i&
            0.325801-0.094442i \\
            			\hline
            $ \alpha$ =-0.4&
            0.283190-0.103629i&
            0.286794-0.103453i&
            0.294792-0.102837i&
            0.320802-0.099054i \\
            \hline
            $ \alpha$ =-0.5&
            0.280176-0.105448i&
            0.283606-0.105465i&
            0.291211-0.105303i&
            0.315738-0.103437i \\
            \hline
            $ \alpha$ =-0.6&
            0.276861-0.107536i&
            0.280116-0.107751i&
            0.287335-0.108048i&
            0.310551-0.107892i \\
            \hline
            $ \alpha$ =-0.7&
            0.273190-0.109971i&
            0.276268-0.110400i&
            0.283107-0.111177i&
            0.305179-0.112655i \\
            \hline
            $ \alpha$ =-0.8&
            0.269110-0.112846i&
            0.272009-0.113508i&
            0.278476-0.114803i&
            0.299590-0.117928i \\
            \hline
            $ \alpha$ =-0.9&
            0.264577-0.116261i&
            0.267298-0.117184i&
            0.273408-0.119049i&
            0.293784-0.123892i \\
            \hline
            $ \alpha$ =-1.0&
            0.259559-0.120331i&
            0.262106-0.121548i&
            0.267885-0.124051i&
            0.287805-0.130715i \\
            \hline
            $ \alpha$ =-1.1&
            0.254040-0.125183i&
            0.256425-0.126733i&
            0.261920-0.129950i&
            0.281750-0.138551i \\            			
            \hline
            $ \alpha$ =-1.2&
            0.248030-0.130956i&
            0.250276-0.132884i&
            0.255560-0.136899i&
            0.275775-0.147526i \\
            \hline
            $ \alpha$ =-1.3&
            0.241574-0.137799i&
            0.243718-0.140151i&
            0.248901-0.145047i&
            0.270101-0.157725i \\            			
            \hline
            $ \alpha$ =-1.4&
            0.234761-0.145861i&
            0.236859-0.148681i&
            0.242093-0.154526i&
            0.265002-0.169172i \\
            \hline
            $ \alpha$ =-1.5&
            0.227734-0.155280i&
            0.229864-0.158602i&
            0.235345-0.165436i&
            0.260801-0.181813i \\
            \hline
            $ \alpha$ =-1.6&
            0.220691-0.166159i&
            0.222951-0.170001i&
            0.228917-0.177815i&
            0.257834-0.195504i \\
			\hline
		\end{tabular}
		\label{tab5.1}
	\end{center}
	Table II. The quasinormal frequencies $\omega$ of the scalar field  calculated by the WKB method with different Gauss-Bonnet constants $\alpha$ and magnetic charges $g$. In the calculation, we let $n=0$ and $l=1$
\end{table}

\begin{table}[htbp]
	\begin{center}
		\begin{tabular}{|c|c|c|c|}
			\hline
			&$l=1,n=0$ &
			$l=2,n=0$&
			$l=2,n=1$\\
			\hline
			$\alpha$=-0.1&
			0.295166-0.098291i&
			0.487380-0.097335i&
			0.467831-0.297406i\\
			\hline
			$ \alpha$ =-0.3&
			0.289742-0.101635i&
			0.479909-0.100419i&
			0.457927-0.307670i\\
			\hline
			$ \alpha$ =-0.5&
			0.283606-0.105465i&
			0.473031-0.103268i&
			0.448363-0.317433i\\
			\hline
			$ \alpha$ =-0.7&
			0.276268-0.110400i&
			0.466670-0.105982i&
			0.439211-0.327056i\\
			\hline
			$ \alpha$ =-0.9&
			0.267298-0.117184i&
			0.460810-0.108623i&
			0.430660-0.336698i\\
			\hline
			$ \alpha$ =-1.1&
			0.256425-0.126733i&
			0.455486-0.111211i&
			0.423009-0.346310i\\            			
			\hline
			$ \alpha$ =-1.3&
			0.243718-0.140151i&
			0.450774-0.113736i&
		    0.416652-0.355644i\\            			
			\hline
			$ \alpha$ =-1.5&
			0.229864-0.158602i&
			0.446782-0.116153i&
			0.412041-0.364258i\\
			\hline
		\end{tabular}
		\label{tab5.1}
	\end{center}
	Table III. The quasinormal frequencies $\omega$ of the scalar field  calculated by the WKB method with different Gauss-Bonnet constants $\alpha$ and multiple numbers $l$. In the calculation, we let $g=0.3$
\end{table}

 From Table II, III , Figure 1 and 2, we see that the dependence of the QNMs and the parameters. For example, we observe that the real part of the QNMs increases when the values of the Gauss-Bonnet coupling constant $\alpha$ increases for the fixed magnetic charge $g$ and same multiple numbers $l$, while the imaginary part of the QNMs decreasing. We see a similar effect of $g$, namely, when $g$ increasing, the real part of QNMs increases and the imaginary part decreases.
 
 \begin{figure}[htbp]
 	\centering
 	\includegraphics[width=.8\textwidth]{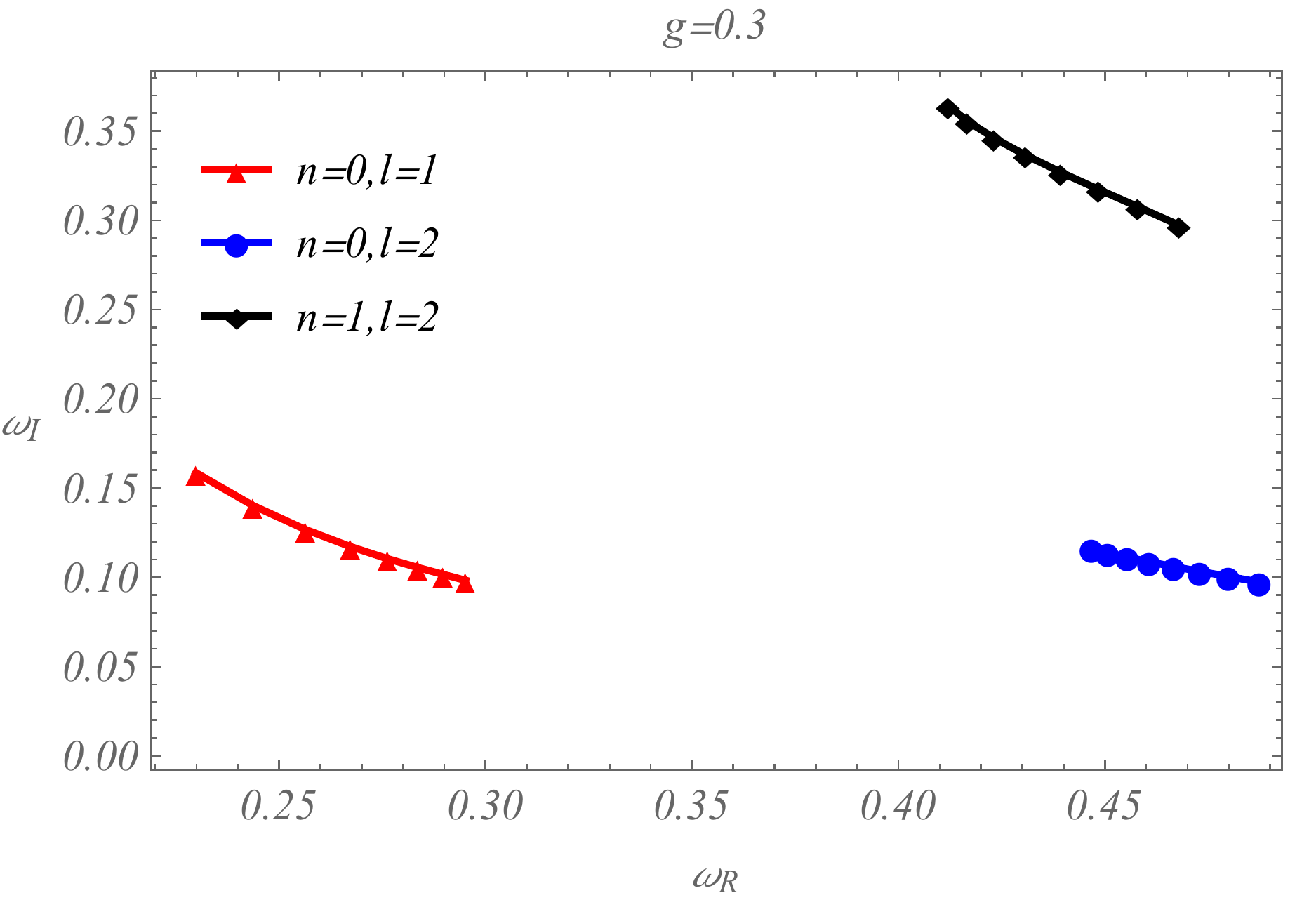}
 	\caption{Dependence of the QNMs on $g$ and $\alpha$}
 \end{figure}
 
 \begin{figure}[htbp]
 	\centering
 	\includegraphics[width=.8\textwidth]{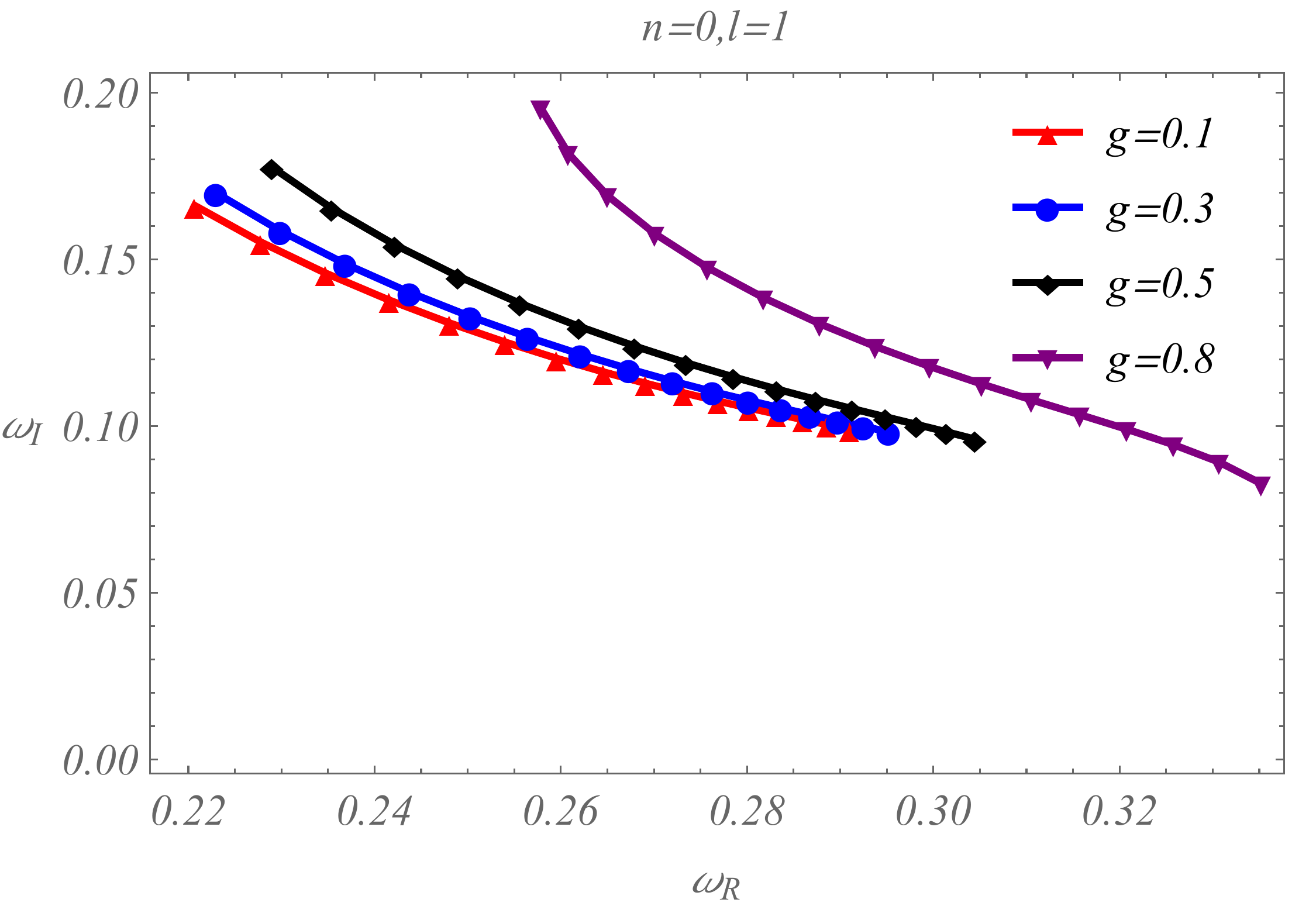}
 	\caption{Dependence of the QNMs on $\alpha$ and $ n, l$}
 \end{figure}
 
 \begin{table}[htbp]
 	\begin{center}
 		\begin{tabular}{|c|c|c|c|}
 			\hline
 			&$\alpha=-0.1$&
 			$\alpha=-0.5$&
 			$\alpha=-0.9$\\
 			\hline
 			$ -\frac{i}{2} \lambda $&
 			$-0.0975136i$&
 			$-0.102257i$&
 			$-0.106043i$\\
 			\hline
 			$R_{sh}$&
 			5.22557&
 			5.36589&
 			5.49105\\
 			\hline
 			$ l=1$&
 			0.291025-0.0990727i&
 			0.280176-0.105448i&
 			0.264577-0.116261i\\
 			\hline
 			RD&
 			52.077{\%}&
 			50.339{\%}&
 			45.281{\%}\\
 			\hline
 			$ l=2$&
 			0.480805-0.0981326i&
 			0.467507-0.103509i&
 			0.45601-0.108521i\\
 			\hline
 			RD&
 			25.624{\%}&
 			25.349{\%}&
 			25.199{\%}\\
 			\hline
 			$l=5$&
 			1.05361-0.097645i&
 			1.02588-0.120502i&
 			1.00236-0.106446i\\
 			\hline
 			RD&
 			10.114{\%}&
 			10.095{\%}&
 			10.08{\%}\\
 			\hline
 			$ l=10$&
 			2.00992-0.0975498i&
 			1.95728-0.102324i&
 			1.9126-0.10615i\\
 			\hline
 			RD&
 			5.03{\%}&
 			5.025{\%}&
 			5.022{\%}\\
 			\hline
 			$ l=20$&
 			3.92331-0.0975231i&
 			3.82067-0.102275i&
 			3.73355-0.106071i\\
 			\hline
 			RD&
 			2.5075{\%}&
 			2.5065{\%}&
 			2.5055{\%}\\
 			\hline
 			$ l=50$&
 			9.66413-0.0975151i&
 			9.41139-0.10226i&
 			9.19687-0.106048i\\
 			\hline
 			RD&
 			1.0012{\%}&
 			1.001{\%}&
 			1.001{\%}\\
 			\hline
 			$ l=100$&
 			19.2324-0.097514i&
 			18.7295-0.102258i&
 			18.3026-0.106044i\\
 			\hline
 			RD&
 			0.5{\%}&
 			0.5{\%}&
 			0.5{\%}\\
 			\hline
 		\end{tabular}
 		\label{tab5.1}
 	\end{center}
 	Table IV.The QNMs derived by the WKB method and shadow radius and the relative deviations. In the calculation, we let $n=0$ and $g=0.1$.
 \end{table}

\begin{table}[htbp]
	\begin{center}
		\begin{tabular}{|c|c|c|c|}
			\hline
			&
			$\alpha=-0.1$&
			$\alpha=-0.5$&
			$\alpha=-0.9$ \\
			\hline
			$ -\frac{i}{2} \lambda $&
			$-0.0967442i$&
			$-0.102024i$&
			$-0.106122i$ \\
			\hline
			$R_{sh}$&
			5.15674&
			5.30551&
			5.43670\\
			\hline
			$ l=1$&
			0.295166-0.0982906i&
			0.283606-0.105465i&
			0.267298-0.117184i\\
			\hline
			RD&
			25.665{\%}&
			25.484{\%}&
			25.264{\%}\\
			\hline
			$ l=2$&
			0.48738-0.0973345i&
			0.473031-0.103268i&
			0.46081-0.108623i\\
			\hline
			RD&
			52.209{\%}&
			50.467{\%}&
			45.322{\%}\\
			\hline
			$l=5$&
			1.06774-0.096868i&
			1.03764-0.102266i&
			1.01246-0.106527i\\
			\hline
			RD&
			10.121{\%}&
			10.104{\%}&
			10.088{\%}\\
			\hline
			$ l=10$&
			2.03678-0.0967783i&
			1.9796-0.10209i&
			1.93176-0.10623i\\
			\hline
			RD&
			5.031{\%}&
			5.028{\%}&
			5.024{\%}\\
			\hline
			$ l=20$&
			3.97569-0.0967531i&
			3.86418-0.102042i&
			3.7709-0.10615i\\
			\hline
			RD&
			2.508{\%}&
			2.507{\%}&
			2.5065{\%}\\
			\hline
			$ l=50$&
			9.79313-0.0967456i&
			9.51852-0.102027i&
			9.28882-0.106127i\\
			\hline
			RD&
			1.0012{\%}&
			1.0012{\%}&
			1.001{\%}\\
			\hline
			$ l=100$&
			19.4891-0.0967445i&
			18.9426-0.102025i&
			18.4855-0.106123i\\
			\hline
			RD&
			0.5{\%}&
			0.5{\%}&
			0.5{\%}\\
			\hline
		\end{tabular}
		\label{tab5.1}
	\end{center}
	Table V. The QNMs derived by the WKB method and shadow radius and the relative deviations. In the calculation, we let $n=0$ and $g=0.3$.
\end{table}
 
\par
There is a close connection between null geodesics and QNMs. It was proven in \cite{VCMBWZ} that real part of QNMs in the eikonal limit is related to the angular velocity of the null geodesic $\Omega_c$, and the imaginary part is related to the Lyapunov exponent $\lambda$ that determines the instability time scale of the orbit. An analytic formula is given by
\begin{equation}
	\omega_{QNM}=\Omega_c l -i(n+1/2)|\lambda|,
	\label{eq3.7}
\end{equation} 
where $n$ is the overtone number and $l$ is the angular momentum of the perturbation. The Lyapunov exponent of the photon sphere can be given as
\begin{equation}
	\lambda=\sqrt{\frac{f(r_{ps})(2f(r_{ps})-r^2_{ps}f''(r_{ps}))}{2r^2_{ps}}}.
	\label{eq3.8}
\end{equation}
Further, Stefanov $\emph{et al.}$ \cite{RAKAFZ} also found such relation between the geodesics and QNMs only valid at high multipole numbers. Jusufi \cite{KJUS1,KJUS2} found a simple relation between the real part of the QNMs and the shadow raduis in the eikonal limit with large values of the $l$
\begin{equation}
	\omega_R=\lim_{l >>1 }\frac{l}{R_{sh}}.
	\label{eq3.9}
\end{equation}
But this correspondence is not guaranteed for any type of filed, for example for gravitational fields in the Einstein-Lovelock theory \cite{RAKZS}. For the four-dimensional charged Einstein-Gauss-Bonnet black holes, Chen $\emph{et al.}$ \cite{CDY} verified such correspondence in the case of scalar filed. Therefore here we gave a relation between the QNMs and shadow radius in the eikonal limit as
\begin{equation}
	\omega=\lim_{l \gg 1 }\frac{l}{R_{sh}}-i(n+\frac{1}{2})\lambda.
	\label{eq3.10}
\end{equation}  

Using above equations and the WKB method, we got the Table IV and V. In these tables, $R_{sh}$ denotes the radius of the shadow radius, $\omega$ denotes the QNMs which calculated by the WKB method. We listed $R_{sh}$ and $\omega$ with different values of $l$ and $\alpha$, RD represent the relative deviations of the product of the real parts of the QNMs $\omega$ and  $R_{sh}$ from the multiple number  $l$. First, from these Tables, we see that for real part of the QNMs, the relative deviations is huge for low values of the $l$, while decreasing as the values of $l$ increasing, for $l=100$ the relative deviation arrives at $0.5 {\%}$, therefore we can conclude the validness of Eq. (\ref{eq3.9}). Secondly, the imaginary part of QNMs calculated from WKB method fits enough well at the low values of $l$ with the value calculated by $-i \lambda /2$. As the values of the $l$ increases, the imaginary part of QNMs is more and more accurate with respect to $-i \lambda /2$. For example, for $\alpha = -0.1, g=0.1, l=100 $, the difference between them is $0.00003i$. If we calculated the relative deviation in this case, the result would be $0.00031 {\%}$. Therefore Eq. (\ref{eq3.10}) works well for both the real part and the imaginary part of the QNMs in the scalar perturbation.

\section{CONCLUSION}
In this paper, we investigated the shadow radius and QNMs of the four-dimensional magnetically charged Einstein-Gauss-Bonnet Bardeen black hole and pointed out a simple connection between them in the eikonal limit related by Eq. (\ref{eq3.10}), in which the real parts of this relation are accurate only for large values of $l$. Note that, even in the limit $l \gg 1$ this
correspondence is not guaranteed for gravitational fields. We have performed detailed analyses of QNMs for a massless scalar in this spacetime background. Using the sixth-order WKB approximation, we found that the QNM frequencies of this black holes strongly depend upon the Gauss-Bonnet coupling constant $\alpha$ and the magnetic charge $g$. This investigation reveals a potential relationship between the black hole shadows and gravitational waves.

\end{document}